\documentclass{article}

\usepackage[english]{babel}

\usepackage[a4paper,top=2cm,bottom=2cm,left=3cm,right=3cm,marginparwidth=1.75cm]{geometry}

\usepackage[backend=bibtex, sorting=none, style=numeric-comp ]{biblatex} 
\appto{\bibsetup}{\raggedright}
\usepackage{amsmath}
\usepackage{graphicx}

\title{Effects of a surrounding environment during the coalescence of AuPd nanoparticles}
\author{Sofia Zinzani,\textit{$^{a}$} and Francesca Baletto \textit{$^{a,b}$}
\\
\textit{$^{a}$~Universita' di Milano, Dipartimento di Fisica, Via Celoria 16, Milano I-20133;} \\
francesca.baletto@unimi.it\\
\textit{$^{b}$~Physics Department, King's College London, WC2R 2LS, UK}
}

\begin{document}
\maketitle

\begin{abstract}
It is far well accepted that the morphology of nanoparticles and nanoalloys is of paramount importance to understand their properties. Furthemore, the morphology depends on the growth mechanism with coalescence generally accepted as one the  most common mechanisms both in liquid and in the gas phase. Coalescence refers when two existing seeds collide and aggregate into a larger object. It is expected that the resulting aggregate shows a compact, often spherical structure, although strongly out of the equilibrium, referring to its global minimum. While the coalescence of liquid droplet is widely studied, the first stages of the coalescence between nanoseeds has attracted less interest, although important as multiple aggregation can take place.  Here we simulate the coalescence of Au and Pd seeds by the Molecular Dynamics method, comparing the initial stage of the coalescence in vacuum and when there is an interacting surrounding around them. We show that the surface chemical composition of the resulting aggregate depend on the environment as well as the overall morphology.
\end{abstract}

\section{Introduction}
Interest in nanoparticles (NPs) never fades since the Granqvist's work and the Japanese project Erato \cite{Granqvist}. A nanoparticle is a discrete object with all three dimensions less than 100 nm. NPs attracted such great interest for their na\"ive properties and they have been often considered a new phase of matter. NPs find application in several technological fields as the building blocks of devices as sensors, energy harvesting devices, catalysts, and random-assembled neuromorphic systems \cite{MELINON1995, Perez1999}. 
Among the class of nanoparticles, metal nanoparticles (MNPs) and metallic nanoalloys (NAs) - referring to MNPs made of two or more metals- possess particular chemophysical properties. Furthermore, the physical properties of MNPs can be tuned from their morphology -meant as a combination of their size, shape, chemical composition, chemical ordering. Such property-morphology dependence paves the way to the design of nanomaterials to satisfy our needs \cite{BookRoy}.
At the core of such rational design process, we need to understand how MNPs can form and agglomerate (self-assembly). 
Numerical methods can shed lights on the atomistic processes and the timescale of the formation process complementary to experimental techniques \cite{Baletto2019, Grammatikopoulos2019}.
Generally speaking, the synthesis of MNPs follows three possible paths. They can grow one-by-one, meaning by the additional deposition of atoms over an existing core; through the sintering of two or more individual seeds; via the annealing of a melted seed.
Both from the coalescence and the one-by-one growth, the surface growth is driven by adatom diffusion and small island formation, which can be altered by the surroundings.
Indeed, several studies report on the coalescence of Au-based nanoparticles \cite{Nelli2021Nanoscale} but only in the perfect vacuum.
 In the following, we focus on this system because Au-Pd NAs can exhibit superior catalytic activity and enhanced selectivity as compared to their monometallic counterparts due to synergistic effects. Au-Pd nanoalloys have been widely investigated in both heterogeneous thermal catalysis and electrocatalysis \cite{doi:10.1021/acs.chemrev.8b00696}, including industrial processes including hydrogen peroxide synthesis from H$_2$ and O$_2$, alcohol oxidation, vinyl acetate monomer synthesis, and formic acid de-hydrogenation. The rational of combining Au with Pd is due to the plasmonic and versatile character of Au-NPs with the storng catalytic flavour of Pd-NPs.
Several previous studies explore the complex energy landscape of Au-Pd showing the tendency to form Au-rich surfaces, ball (Pd)-cup (Au) ordering at small sizes, while at larger sizes a more alloyed inner core as expected looking at the Au-Pd phase diagram. We should note that often both global searches as well as growth studies are only in the perfect vacuum achievable in numerical studies \cite{Nelli2021JPC}.
To explore the catalytic activity of NAs, the morphology and in particular the surface composition Au-Pd NAs should be revealed changing the surrounding. 
Our work aims to investigate and compare the kinetics of the coalescence of Au-Pd NAs over few tens of ns. We purpose to elucidate the initial phases of the sintering process in the vacuum and in a plethora of interacting environments. using the Huerto-Cortes' formalism to mimick the presence of a uniform environment around the nanoparticle, we clearly show that the surface chemical composition can be modified depending on the surroundings and that the environment can also modify the timescale of the sintering process.

\section*{Methodology}
We perform classical molecular dynamics simulations, using the open-source package \texttt{LoDiS} \cite{site:LoDiS}, at fixed temperature using an Andersen thermostat with a frequency of 10$^{11}$Hz to generate our trajectories.
We average our results over independent simulations for every system and temperature considered in order to collect statistics and ensure a greater reliability. 
We fix the timescale at 10 ns, as we are interested in the first stages of the formation process and when likely major structural changes occur, in such a way that photo-catalytic properties might be strongly affected. 
For our simulation we used both a metal-metal potential and a metal-environment potential.

\subsection{Metal-metal potential}
The metal-metal interaction is modeled according to the Gupta or Rosato-Guillope-Legrande formulation, \cite{art:potrosato}, which is derived as part of the second moment approximation (SMA) in the tight-binding (TB) model.
The contribution of the band to the cohesive energy, derived according to the second moment approximation of tight binding, is
$$
E_{b}^i (r_{ij})=\sqrt{\sum_{j \neq i}^{n_v} \xi_{a b}^2 e^{-2 q_{a b}\left(\frac{r_{i j}}{r_{a b}^0}-1\right)}}
$$

where $r_{ab}^0$ is the cut-off radius, defined as the arithmetic mean of the bulk nearest neighbour distances for atoms of type a and b in the bimetallic case and $r_{ij}$ is the atomic pair distances.
The sum is up to the number of atoms $n_v$ within an appropriate cut-off distance from atom i, where a and b
refer to the chemical species of the two atoms.

The repulsive contribution is well described by a pair interaction such as the Born-Mayer potential:
$$
E_{r}^i (r_{ij})=\sum_{j \neq i}^{n_v} A_{a b} e^{-p_{a b}\left(\frac{r_{i j}}{r_{a b}^0}-1\right)}
$$
Considering then the potential as a whole, we will have:
$$
E_{T B S M A}^i (r_{ij})=\sum_{j \neq i}^{n_v} A_{a b} e^{-p_{a b}\left(\frac{r_{i j}}{r_{a b}^0}- 1\right)}-\sqrt{\sum_{j \neq i}^{n_v} \xi_{a b}^2 e^{-2 q_{a b}{\left(\frac{r_{i j}}{r_{a b}^0}-1\right)}}}
$$
The parameters $A_{a b}$, $\xi_{a b}$ , are determined to reproduce the experimental values of the cohesive energy, while $p_{a b}$ , $q_{a b}$ and their product tune the ’stickyness’ of the potential.

The parameterization in the case of our bimetallic (gold-palladium) system is given by: 

\begin{center}
    \begin{tabular}{ c|c|c|c|c } 
    & $p$ & $q$ & $A$ [eV] & $\xi$ [eV]\\
    \hline
    Au-Au & 10.139 & 4.033 & 0.209570656 & 1.8152764 \\ 
    \hline
    Pd-Pd & 11.0 & 3.794 & 0.1715 & 1.7019 \\
    \hline
    Au-Pd &10.543 & 3.8862 & 0.1897 & 1.7536 \\ 
    \end{tabular}
\end{center}

\subsection{Metal-environment potential}
Metal-environment interactions are obtained as the collective of atomic contributions dependent on the coordination number ($CN_i$) using Huerto-Cortes, Goniakowski, Noguera formalism \cite{Cortes-Huerto}
\begin{equation}
    E^{M-E}_i= -\eta (CN_{bulk}-CN_i)^{\rho}
\label{eq:potimplicit}
\end{equation}
The difference between the bulk coordination number $CN_{bulk}$ and $CN_i$ is analogous to the number of absent bonds with respect to bulk. For FCC metals, $CN_{bulk} = 12$. 
The nature of the interaction and its strength is determined for each chemical species of the system (Au and Pd).
The nature of the interaction is determined by $\rho$:
\begin{itemize}
    \item $\rho = 1$, pairwise interaction
    \item $\rho < 1$, covalent bonding
    \item $\rho > 1$, strongly interacting environments
\end{itemize}

The $\eta$ parameter fixes the interaction strength. Different $\rho$ and $\eta$ parameters set the ratio between the surface energies of low Miller index terminations, thus introducing a tunable parameter to favour an architecture or another.

\subsection{Descriptors for characterization} 
We conduct the analysis of the classical MD trajectories by means of the \texttt{Sapphire} \cite{art:sapphire}.
Below, we report the selected quantities to characterise the evolution of the coalescence of AuPd nanoalloys. 

\subsubsection{Pair Distance Distribution Function}

\begin{figure}[ht!]
    \centering
    \includegraphics[width=.5\textwidth]{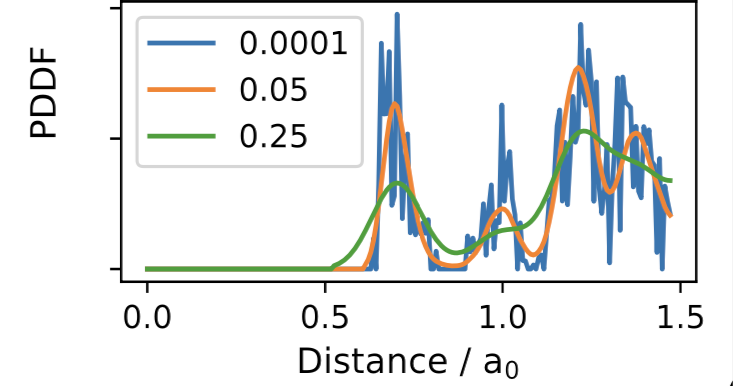}
    \caption{Dependence on the bandwidth, $h$ in Eq.\ref{eq:PDDF}, orange (0.05), blue (0.0001) and (0.25) green with respect to the bulk lattice parameter $a_0$. Data, taken with permission form RM Jones' thesis (to be deposited at King's College London library), are taken for an alloyed AuPt NA of 1415 atoms with an Ih motif and a 4:1 AuPt ratio.}
    \label{fig:fromRob}
\end{figure}

The distribution of pair-atomic distances (pair-distance distribution function, PDDF) is a crucial quantity to characterise the geometry and chemical ordering of a MNP.
We define the \textit{pair-distance} as the distance $d_{i j}$ between atoms $i$ and $j$. Given the Cartesian coordinates of the atom $i$ and $j$, $(x_i,y_i,z_i)$ and $(x_j,y_j,z_j)$ respectively,
\begin{equation}
d_{i j}=\sqrt{\left(x_i-x_j\right)^2+\left(y_i-y_j\right)^2+\left(z_i-z_j\right)^2} \mbox{~~~.}
\end{equation}
From the definition of the pair distance, its distribution function can also be studied.
In \texttt{Sapphire}, the \textit{Pair Distance Distribution Function} (PDDF) is calculated via a Kernel function, $K\left(d_{i j}, d ; h\right)$ over the $d_{i j}$ variable, constructed from \textit{n} observations:
\begin{equation}
\operatorname{PDDF}\left[K\left(d_{i j}, d ; h\right)\right]=\frac{1}{N h} \sum_i^N \sum_{j \neq i} K\left(\frac{d_{i j}-d}{h}\right) \mbox{~~~.}
\label{eq:PDDF}
\end{equation}
The parameter $h$ is the bandwidth that defines the tightness of the kernel function. Setting the bandwidth is a delicate step, as described in Refs. \cite{Delgado2021, art:sapphire}. In our analysis, we select $h= 0.05$ of the bulk lattice as it has been proven to be sufficiently broad to smooth the sharp Dirac peaks from having a finite sample, and to sufficiently resolve key features \cite{ThesisRobert}, as shown in Figure \ref{fig:fromRob}.

The PDDF helps our analysis in the following way.
\begin{itemize}
\item The first PDDF minimum is useful to determine the cut-off distance to label the nearest neighbours. Hence s enables the definition of an adjacency matrix and hence of all coordination properties, including common neighbour analysis.
\item The presence of the second peak around the bulk lattice parameter ($a_{0}$) suggests the appearance of a structural order. Indeed, a robust definition of "amorphous/molten-like" shape arises when the peak in correspondence of the bulk lattice is missing. In absence of a PDDF peak at $a_0$ and combination with a radial expansion, the MNP structure can be defined as melted \cite{Delgado2021}.
\item The maximum of the PDDF provides an estimate of the size of nanoparticle and it should be compared with the radius of gyration, see later, to provide a clue about the sphericization process (or neck sintering stages). So, we define the \textit{half maximum pair distance function}: $R_{max}$, defined as the half of the maximum pair distance between all the pair of atoms of the nanoparticle.
\end{itemize}

\subsubsection{Radius of gyration}

The radius of gyration of a body is defined as the radial distance from a point having a moment of inertia equal to the effective mass distribution of the body if the total mass of the body were concentrated at that point.
To facilitate the analysis of the radial distribution function, one can refer to the radius of gyration of a nanoparticle, i.e. the measure of the extent by which atoms in the system are spread out away of the center of mass ($\mathrm{CoM}_{w}$) of the whole nanoparticle.

\begin{equation}
G=\sqrt{\frac{1}{N} \sum_{i=1}^N\left(r_i-\mathrm{CoM}_{w}\right)^2} .
\end{equation}
Conversely, contrasting the radius of gyration of the two A and B chemical species present in a bimetallic nanoalloy,
\begin{equation}
\begin{aligned}
G^A &=\sqrt{\frac{1}{N} \sum_{i \in A}^N\left(r_i-\mathrm{CoM}_{w}\right)^2}, \\
G^B &=\sqrt{\frac{1}{N} \sum_{i \in B}^N\left(r_i-\mathrm{CoM}_{w}\right)^2}
\end{aligned}
\end{equation}
will yield precise information on the chemical ordering of the nanoparticle itself: similar radii of gyration for the two chemical species will be observed for mixed or Janus ordering, while shelled-structures will be characterized by $G(r)^A$ and $G(r)^B$ of markedly different values.

\subsubsection{Nominal and Generalized Coordination Number}

With a focus on catalysis, the coordination of an adsorption site has been often used as a descriptor to rationalize its activity.
A simple chemical intuition suggests that low-coordinated atoms are more likely to form chemical bonds than highly-coordinated ones.
To understand the coordination of atoms it is necessary to define which are the nearest neighbor. 

For this purpose, the \textit{adjacency matrix $A(r_{i j})$}, defined as 1 if the distance between atoms is less than the separation radius and 0 if it is greater, was introduced. 
\begin{equation}
\begin{gathered}
A\left(r_{i j}\right)= \begin{cases}1 & \text { if } r_{i j} \leq R_{cut} \\
0 & \text { if } r_{i j}>R_{cut}\end{cases}
\end{gathered}
\end{equation}
Operationally, the cut-off radius is determined by the first minimum of the PDDF.

To figure out whether an atom is low- or high-coordination, we can introduce the coordination number CN.
Considering a cut-off region, described by a sphere of radius equal to the proximity distance of the bulk and centered around the atom, we can enumerate the other atoms that fall within that region, thus defining the total coordination number of the nanoparticle:
\begin{equation}
\begin{gathered}
C N_{t o t}=\sum_{i \neq j} A\left(r_{i j}\right) \\
\end{gathered}
\end{equation}
where $A(r_{i j})$ is the Adjacency Matrix.

Starting from the definition of the CN, we can derive atop generalized coordination number aGCN. The coordination number (CN) is the number of the nearest neighbor atoms, whereas GCN includes information about the second nearest neighbor atoms to the atoms at the adsorption site. 
By expanding GCN, a clear and rapid analysis of adsorption and related properties can be realized.
The generalized coordination of a set of atoms i, $GCN_i$, is calculated as the sum over the coordination number $CN_j$ of their j neighbours, normalized with respect to the set of atoms bulk coordination:
\begin{equation}
GCN_{i}=\sum_{j} \dfrac{CN_{j}}{CN_{max}}
\end{equation}
with $CN_{max}$ set to 12 as this is the coordination of an FCC atom in the bulk. In general, $CN_{max}$ could be set to 18 for coordination of bridge and set to 24 for hollow sites. With $CN_{max}$ set to 12, we can talk about Atop Generalised Coordination Number: \textit{aGCN}. This is a weighted coordination where the local environment of the atom i with the coordination $CN(j)$ of the j-neighbour of the atom i, divided by the maximum coordination in the bulk.

In the case of nanoalloys, we can easily separate the local environment of each atom between homo-pairs and hetero-pairs. Counting the $A-A$, $N B_{A A}$, $B-B$ $N B_{B B}$, and A-B pairs, $N B_{A B}$ enables to evaluate the mixing parameter $\mu$. The latter is a useful parameter for a fast characterization of the NAs chemical ordering :
\begin{equation}
\mu=\frac{N B_{A A}+N B_{B B}-N B_{A B}}{N B_{A A}+N B_{B B}+N B_{A B}}
\end{equation}
where $\mu$ tends to $-1$ when the NA is fully alloyed, and to $+1$ when there is a complete phase separation.

\subsubsection{Common neighbour signatures distribution}

To classify the geometrical environments of core and surface atoms, we can introduce a more complex characterization procedures: common neighbour analysis (CNA).
CNA characterizes pair of nearest neighbours depending on the local connectivity of their shared neighbours. CNA signatures are of the form (r,s,t) such that r is the number of nearest-neighbours common to both atoms in the pair; s is the number of bonds between shared neighbours and t is the longest chain which can be made from bonding s atoms if they are nearest neighbours.
While an individual CNA signature describes the local environment of a pair of neighbours, the estimate of the percentage of how many pairs contributes to selected CNA signatures provides information on the overall nanoparticle’s structure, allowing a fast classification into expected geometrical family, as icosahedral, decahedral, or FCC-like.

\subsubsection{CNA-Patterns}

While each CNA signature refers to pair of atoms and their average is a property of the whole nanoparticle, the local environment of each atom can be translate into the {\it CNAP}, referring to the pattern of CNA signatures that the atom contributes to.
Per a given atom $i$, we list all the CNA signature $(r,s,t)$ it contributes to, and calculate the frequency (times) such signature occur $f$. The CNAP is then a string $[f_1(r,s,t)_1; f_2(r,s,t)_2; ... ]$ containing all the $k-th$ CNA- signatures $(r,s,t)_k$ the atom is in. The frequency of each signature correates with the coordiantion of the atom $i$. Hence, for atom residing in the inner part of a MNP, we expect $\sum_k f_k = 12$, while atoms at the surface display $\sum_k f_k < 12$.
 In principle, an object with $N$ atoms will have $N$ different CNAP. However, the number of different CNAP decreases when the symmetry of the shape increases. For the same token, the most disordered morphology has that each atom has its own CNAP.
Generally speaking, an automatic classification of NAs and MNPs into the three main morphology families as FCC-like, Dh, and Ih can be done on the basis of CNAP. The [10,(422), 2(555)]; [4,(311), 2,(322), 2,(422)] occur in both Dh and Ih, but the latter is expected to have one [12,(555)] pattern referring to the Ih-centre. Line defects, and plane dislocations in an FCC-like morphology arise into a mixture of $[(l,(422)); (k,(421))]$ with $l + k = 12$.

\subsubsection{Surface identification}

Before illustrating the quantities we have implemented to analyze the surface, it is necessary to define what the surface is.
In particular, we identify, by their atop generalized coordination number, as surface atoms those that do not belong to the nucleus.

Once the surface atoms were then identified through \texttt{Sapphire}, defined as atoms with an atomic generalised coordination number less than 10. We introduced the following three quantities for analyzing surface composition trends. Let $N_A^S$ and $N_B^S$ be the number of atoms of chemical species A and B presents at the surface, and $N_A$ and $N_B$ the total number of atoms of type A and B respectively, and $N$ the total number of atoms in the nanoalloys, we considered
\begin{itemize}
    \item the percentages of type A atoms on the surface compared to the total number of atoms on the surface (both type A and type B), and defined \textit{total percentage}:
    \begin{equation}
        P_{A_{S}/N_S}:= \dfrac{N_A^S}{N_A^S+N_B^S}
    \end{equation}
    and
    \begin{equation}
        P_{B_{S}/N_S}:= \dfrac{N_B^S}{N_A^S+N_B^S}
    \end{equation}
    
    \item the percentage of surface atoms of a chemical species compared to the total number of atoms of the same chemical species, and defined that percentage \textit{relative percentage}:
    \begin{equation}
        P_{{A_{S}}/N_{A}}:= \dfrac{N_A^S}{N_A}
    \end{equation}
    and
    \begin{equation}
        P_{{B_{S}}/N_{B}}:= \dfrac{N_B^S}{N_B}
    \end{equation}
\end{itemize}

\subsubsection{Local Atomic Environment}
We define the local heterogeneous atomic environment (LAE) as being the percentage of species B = Pt with LAE neighbours of species A = Au. LAE $=$ 0 indicates that these atoms create no heterogeneous bonds. 1 $ <=$ LAE $<=$ 6 describes a mix environment; while LAE $>=$ 9 suggest that these atoms (Pt) are almost totally encapsulation by species Au atoms. 

\section{Results}

\begin{figure}[ht!]
    \centering
    \includegraphics[width=.3\textwidth]{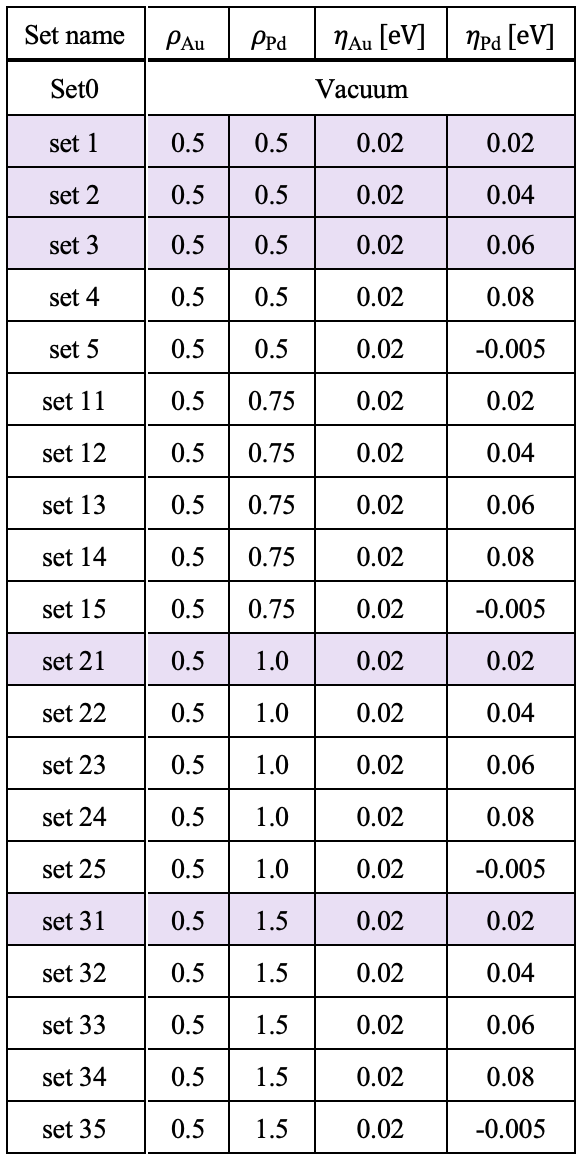}
    \caption{Sets of simulations at 400K and 600K performed for both nanoalloys composed of 55 atoms of gold and 55 of palladium and in light purple for nanoalloys composed of 561 atoms of gold and 561 of palladium at 600K. 
    On the first columns we reported the set-name corresponding to the choosen parameters $\rho_{Au}$, $\rho_{Pd}$, $\eta_{Au}$, $\eta_{Pd}$ given in columns 2,3,4,5 respectively.}
    \label{fig:set}
\end{figure}

We investigate the effect of the environment on the coalescence of nanoparticles choosing one system composed of 55 gold and 55 palladium atoms icosahedral and one of 561 gold and 561 palladium atoms icosahedral.
We emphasise that in this type of investigation, we have kept constant both the ratio between the quantities of the two chemical species (equal to 50\%) and the shape of the seed: icosahedral for both nanoparticles, at both sizes considered.
For both the systems we consider as a reference the coalescence in vacuum of nanoparticles of the same composition. We will call it "set0". We investigated for the same size and shape of nanoparticles the effect of the environment by varying the parameters of metal-environment potential [Eq. \ref{eq:potimplicit}] and temperature. Table \ref{fig:set} reports all the sets considered: while all the value reported has been considered for system composed of 55 gold and 55 palladium atoms, only the sets highlighted in purple has been considered for the system composed of 561 gold and 561 palladium atoms.
Due to the high computational cost of the simulations in implicit environment and considering that we observed that in vacuum the coalescence (neck formation and spherification) process occurs in the first ns, we have choosen to perform simulations for a time of 10 ns for the smaller system and of 20 ns for the bigger.

\subsection{The case of $Au_{55}^{Ih}$ - $Pd_{55}^{Ih}$ }
We illustrate below what was obtained for simulations in vacuum and in the environment for small nanoparticles (55 atoms per chemical species) at both temperatures of 400K and 600K.
For the sake of completeness we emphasise that we performed simulations with the same parameters also at a temperature of 600K. Although for nanoparticles of this size at 600K, gold-melting effects also takes place, whereas the results at 400K should not present this type of bias. 

Each set of simulations are distinguish for the parameters $\rho_{Au}$, $\rho_{Pd}$, $\eta_{Au}$, $\eta_{Pd}$ used in the Huerto-Cortes potential \ref{eq:potimplicit}.

We set the gold - environment interaction to be always covalent-like ($\rho_{Au}=0.5$). On the other, we varied the interaction palladium - environment from covalent $\rho_{Pd}=0.5$ and $\rho_{Pd}=0.75$, pairwise ($\rho_{Pd}=1.0$) and strongly interacting ($\rho_{Pd}=1.5$). The same pairs of $\rho_{Au}$ and $\rho_{Pd}$, are denoted by the same value of the tens digit of the set name. The sets with only units: set1, set2, set3, set4, set5 possess $\rho_{Au}=0.5$ and $\rho_{Pd}=0.5$.
The sets of "ten": set11, set12, set13, set14, set15 possess $\rho_{Au}=0.5$ and $\rho_{Pd}=0.75$. The sets of the "twenty": set21, set22, set23, set24, set25 possess $\rho_{Au}=0.5$ and $\rho_{Pd}=1.0$. The sets of "thirty": set31, set32, set33, set34, set35 possess $\rho_{Au}=0.5$ and $\rho_{Pd}=1.5$

Furthermore, we varied the relative weight $\eta$ of the interaction by assigning it nomenclature corresponding to the variation of units (from 1 to 5) within the set name.
In the sets whose unit is equal to 1 (set1, set11, set21, set31, set41) the environment has same binding toward gold and palladium $\eta_{Au}=\eta_{Pd}=0.02$. Therefore, we kept the binding to gold fixed $\eta_{Au}=0.02$ and varied the binding to palladium. 
In sets ending with 2, 3 and 4 we are increasing the interaction of palladium with the environment respectively $\eta_{Pd}=0.04$, $\eta_{Pd}=0.06$, $\eta_{Pd}=0.08$. While in sets which name end with 5, we have selected a repulsive interaction $\eta_{Pd}=-0.005$.

\subsubsection{Coalescence Au$_{55}$Pd$_{55}$ at 400 K}

We begin by presenting the results for coalescence between a nanoparticle composed of 55 gold atoms and one composed of 55 palladium atoms, both icosahedral in vacuum, at a temperature of 400K.
Analysis of the trajectories shows that the gold and palladium atoms tend to remain segregated in their two components with a chemical ordering similar to Janus, with the formation of some five-symmetry. However, by slicing the nanoparticle, we can see a preference for gold to arrange itself on the surface and palladium to remain in the core. 
We can also see that from the first few nanoseconds, the two nanoparticles tend to form a rather compact object.

From the trajectory analysis in the environment we can observe that after a few nanoseconds, the two nanoparticles of gold and palladium compose a rather compact nanoalloy. 

\begin{figure}[ht!]
    \centering
    \includegraphics[width=2.55cm]{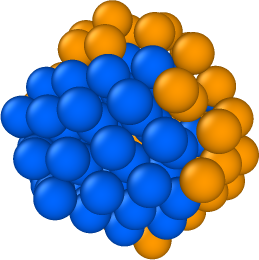}
    \includegraphics[width=2.55cm]{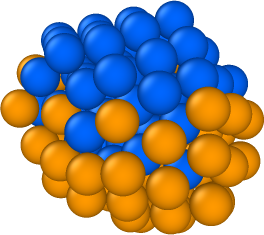}
    \includegraphics[width=2.55cm]{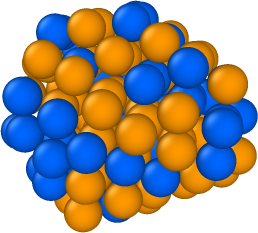}
    \caption{Coalescence ${Au_{55}^{Ih}}$-${Pd_{55}^{Ih}}$ at 400K left panel in vacuum; middle set1; and right panel set34. the single snapshot is taken after 10 ns.}
    \label{fig:snap_55_400K}
\end{figure}

In figure \ref{fig:snap_55_400K} we reported, by way of example, one snapshot for the vacuum (at left), one for the set1 (in center), which serves as a reference for the typical behavior of sets considered, and one for the set34 (right) which possess a particular behavior and serves as a reference also for the set33.
The general tendency is to have the gold spread over the surface in a ball-cup type pattern (Fig. \ref{fig:snap_55_400K}). Only in a few cases, for example set34 (represented in right in Fig. \ref{fig:snap_55_400K}), is there a mixing pattern.
In terms of nanoparticle structure, on the other hand, we frequently find fivefold symmetries, especially for palladium. Sometimes the beginning of the formation of a decahedron is recognizable, however, due to the small number of atoms of both gold and palladium and the presence of both in equal amounts, complete formation cannot be observed in any case.

In following figures we show, the set0 on top, the set1 down on left and set34 on right. Set1 can be considered as a reference for all sets except set34 and set33, which are represented by set34.

\begin{figure}[ht!] 
\centering
\includegraphics[width=4.85cm]{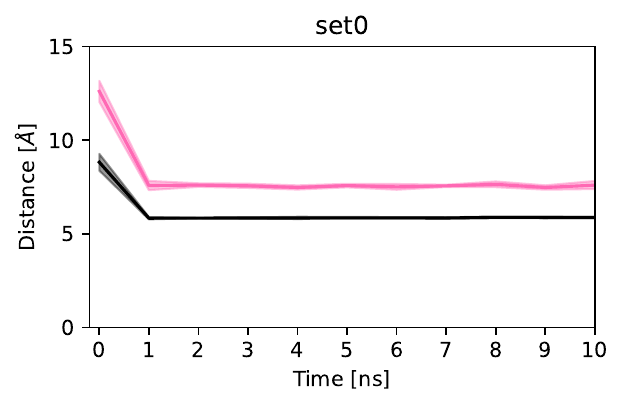}
\includegraphics[width=9.cm]{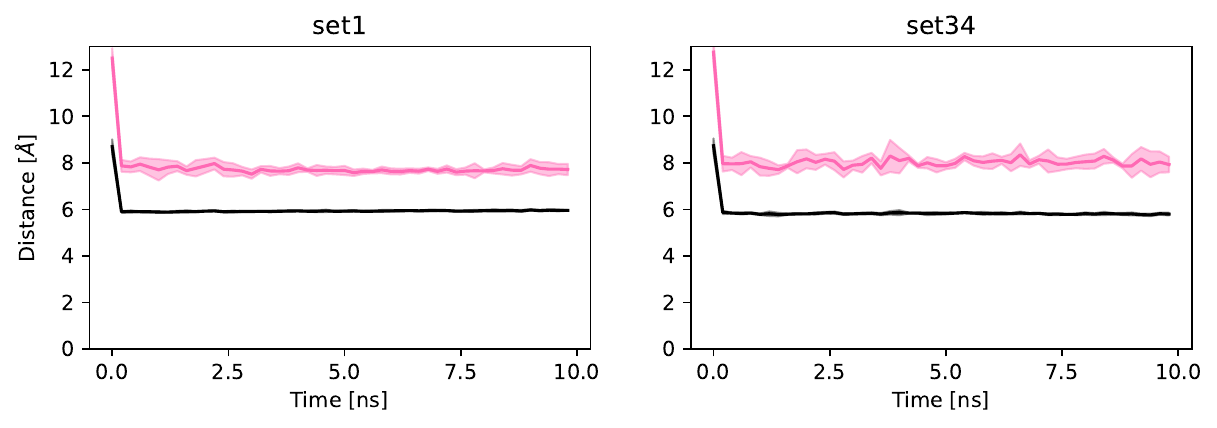}
\caption{Time-varying radius of gyration of the whole cluster (in black) and half maximum pair distance (in pink), for ${Au_{55}^{Ih}}$-${Pd_{55}^{Ih}}$ in vacuum (on the left) and in two sets of the environment (in center and right). }
\label{fig:Union400}
\end{figure}

\paragraph{Kinetic of the '{\it sphericalisation}' process}
From both radius of gyration and maximum pair distance (black and pink lines in Fig. \ref{fig:Union400}), we can observe that either in vacuum and in implicit environment, the structure compacts after a few nanoseconds with a very fast '{\it sphericalisation}' process: in very few ns the cluster becomes a sphere with a radius of gyration of around 6.0 \AA and a radius given by half the maximum distance in pairs of around 8 \AA.
From the radius of gyration we can see that it does not exhibit a tendency to separate again, while calculating its ratio with the half of the maximum pair-distance, and obtaining that it is about $\xi \approx 0.7$ , we can affirm that it does not tend to assume particularly elongated shapes.

\paragraph{Structural characterization}

\begin{figure}[ht!]
    \centering
    \includegraphics[width=4.55cm]{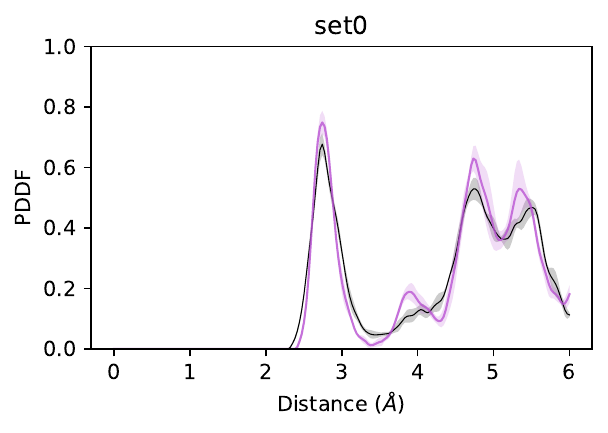}
    \includegraphics[width=9cm]{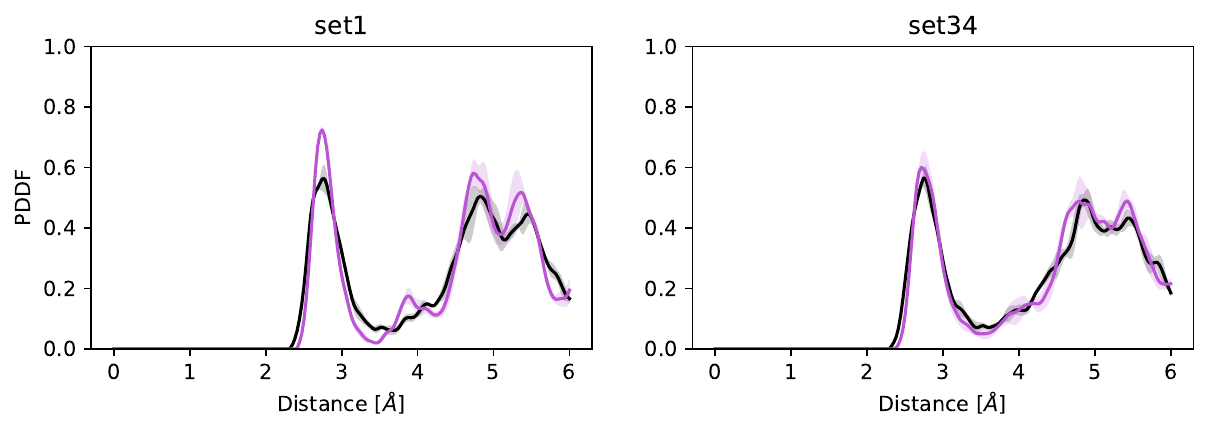}
    \caption{Pair distance distribution function mean of the whole cluster. In black is represented the PDDF at initial time (equal to the contact time) while in purple is represented the PDDF at final time. Final time is equal to 10 ns for all simulations.}
    \label{fig:PDFmean400}
\end{figure}

Considering the set0, we can observe from the pair distance distribution function at the contact time (black line in the figure \ref{fig:PDFmean400}) and at the final time (purple line in the figure \ref{fig:PDFmean400}), that the nanoparticle at the final time presents a more ordered structure. In fact, the peak of the neighbouring seconds, which is completely absent at the initial time, forms at the final time.
The first peak is around 2.9-3 \AA  , while second peak, identified at 3.9-4 \AA .They are both compatible with the gold and the palladium parameters.
Indeed, recall that the first peak (the peak of the first neighbor) is located at 2.88 \AA  for gold and 2.75 \AA  for palladium \cite{Kittel2004}.
We set the gold lattice constant to 4.0782 \AA~  and 3.8907 \AA~ for palladium.

Figure \ref{fig:PDFmean400} shows the pair distance distribution function for the initial and the final snapshot (10ns) for set1 and set34 in environment. 
We can notice that for set1 and all those it represents the final frame has more pronounced peaks. Moreover, while in the initial configuration the second peak is very poorly defined, so much so that it resembles a shoulder rather than a peak itself, at the final time the second peak is well present: thus indicating the presence of a more defined symmetry for the nanoparticle.
An exception occurs for set34, as for all previous quantities, in which the final configuration loses the second peak and the nanoparticle turns out to be less symmetrical than in the initial configuration.
For the purpose of defining the shape of nanoparticles, we analyzed signatures and patterns. 
For set0 we observe, from the analysis of the signatures, that a the absolute values of the (5,5,5) do not vanish, and that there are more (4,2,2) than (4,2,1), this fact leads us to identify a decahedral or icosahedral structure. Adding the information derived from the analysis of the CNAP, we concluded that in the 25\% of cases (one simulation of four) there is a icosahedral shape, while in the other 75\% a decahedral one. This is in agreement with what is observed from the analysis of trajectories.
For the environment, from the analysis of the signature we can observe that although the absolute values of the (5,5,5) are very low, there are significant fluctuations. These testify to the creation and destruction of five-symmetries or five-symmetry points, depending on whether the value of the percentage is exactly zero or not.
Where the percentage is zero, we can detect the presence of an FCC, because of the low value we could assume the presence of either icosahedra or decahedra, the latter being more likely.
We can observe that in all sets there is a high presence of (4,2,2) signatures, this means that in no case is a pure FCC present: in that case (4,2,2) would be zero. Where instead we observe a decrease in (4,2,1) we can expect more decahedral shapes, for example in set2.
We therefore note that classification of the environment can therefore have an effect on the structure: signatures vary between sets.
All structures have many dislocations as (4,2,2) is always present. In many cases we have the presence of (5,5,5), not to such an extent as to justify the presence of icosahedra except with very off-centre axes, whereas it suggests the presence of decahedra.

By analysing the RDFs (not reported but available on request) both in vacuum and in implicit environment, we note an overlapping of the distances at which the atoms of the two chemical species are placed, while observing it in its entirety we do not notice clearly defined icosahedral shell structures.

\begin{figure}[ht!] 
\centering
\includegraphics[width=4.85cm]{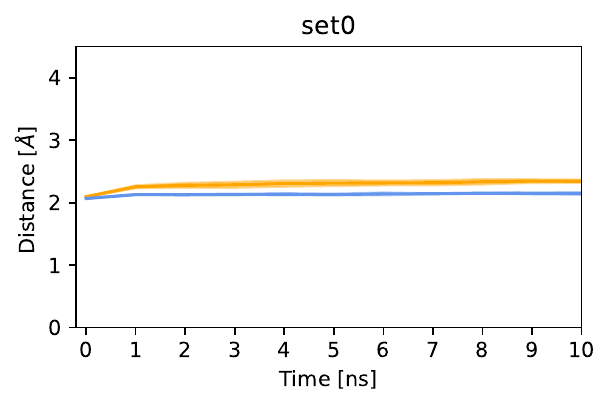}
\includegraphics[width=9.cm]{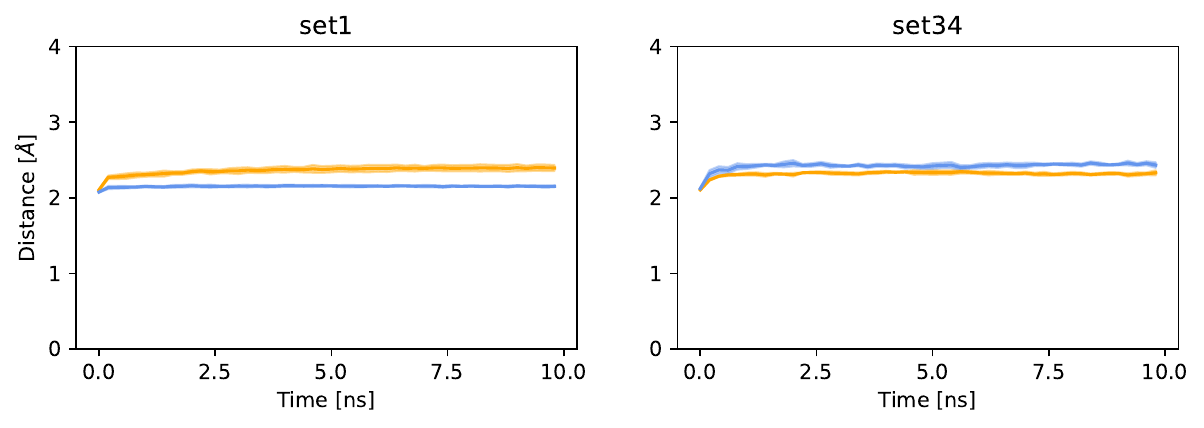}
\caption{Time-varying radius of gyration of the two chemical species: in orange is represented the mean radius of gyration of Au and in blue that of palladium.}
\label{fig:Gyrmean400}
\end{figure}

\paragraph{Chemical ordering and surface composition}
From the separate study of the radius of gyration of gold and palladium (orange and blue lines in Fig. \ref{fig:Gyrmean400}), we can observe that except for set34, the radius of gold is larger than that of palladium, both with values around 2 \AA. This would support the previous hypothesis of palladium being covered by gold. Exception occurs for set34 where the radius of rotation of palladium is greater, as shown in the figure, and there are more oscillations of the maximum distance. 
The trends are similar to set1 for the void, represented in set0.

\begin{figure}[ht!]
    \centering
    \includegraphics[width=4.55cm]{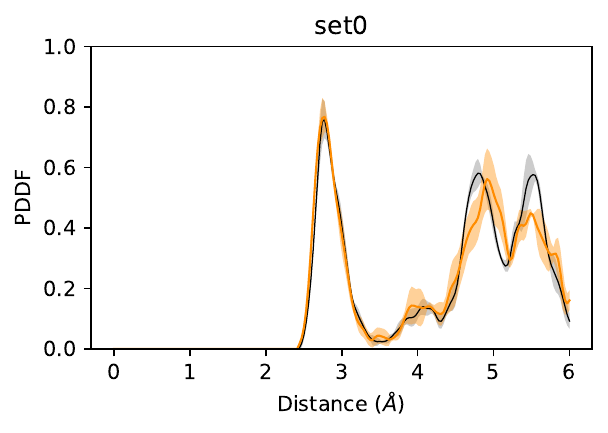}
    \includegraphics[width=9cm]{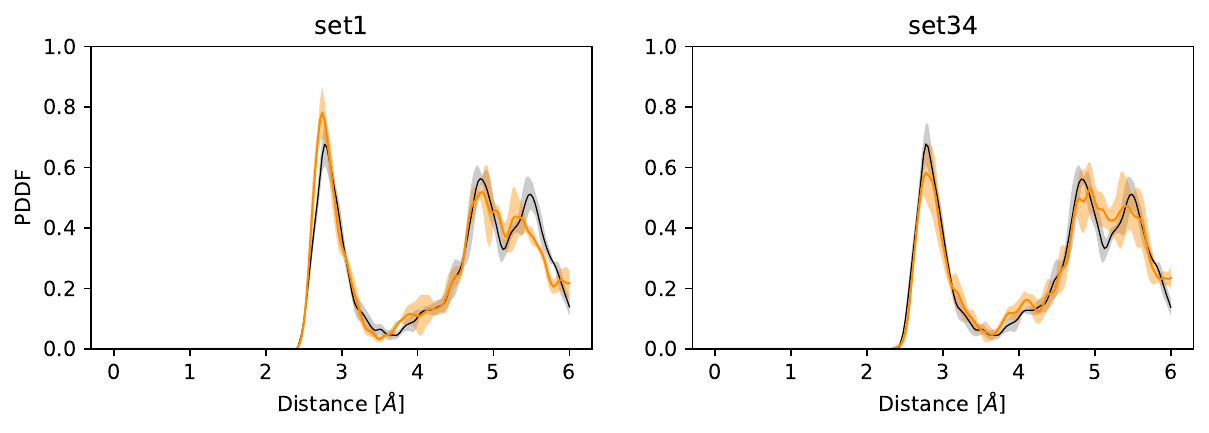}
    \caption{Pair distance distribution function mean of gold. In black is represented the PDDF at initial time (equal to the contact time) while in yellow is represented the PDDF at final time. Final time is equal to 10 ns.
    }
    \label{fig:PDFmeanAu400}
\end{figure}

\begin{figure}[ht!]
    \centering
    \includegraphics[width=4.55cm]{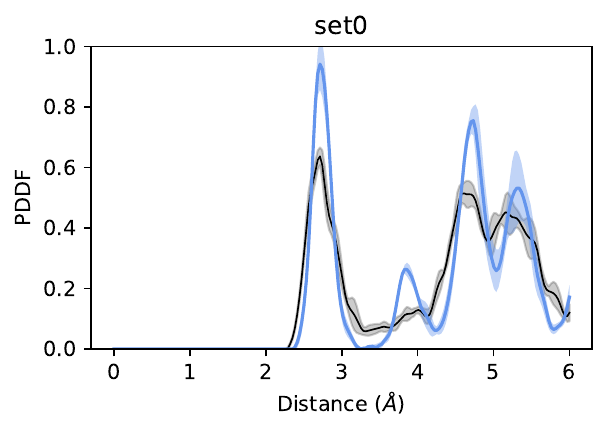}
    \includegraphics[width=9cm]{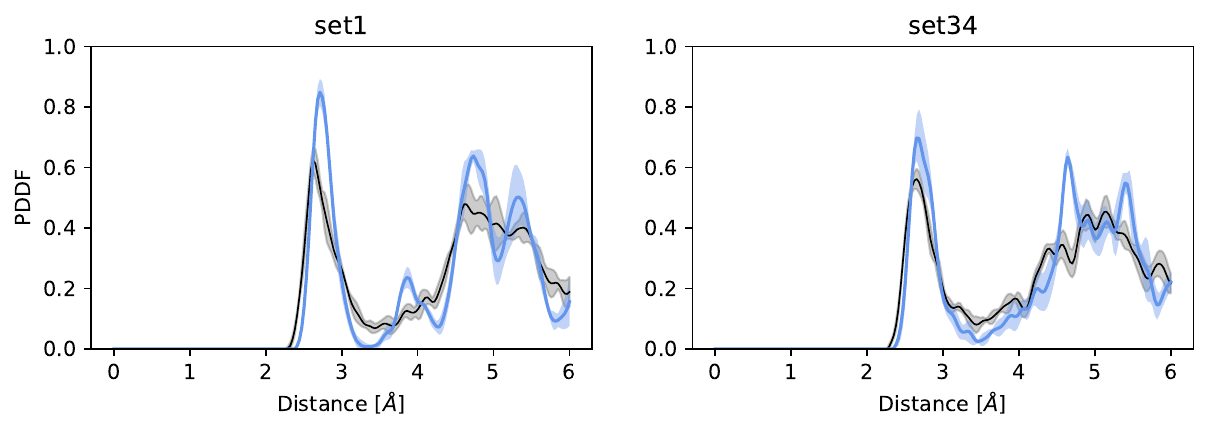}
    \caption{Pair distance distribution function mean of palladium. In black is represented the PDDF at initial time (equal to the contact time) while in blue is represented the PDDF at final time. Final time is equal to 10 ns.}
    \label{fig:PDFmeanPd400}
\end{figure}

Instead, palladium and gold PDDFs for the same sets are shown distinctly in Fig. \ref{fig:PDFmeanAu400} and Fig. \ref{fig:PDFmeanPd400}.
We can observe that for both case of vacuum (set0) and for all the sets in environment except set33 and 34, the palladium-only PDDF (panel 3A) has a well pronounced peak corresponding to the second near neighbour peak at the final time, while gold-only PDDF (panel 4A) has less pronounced/absent peaks. Differently, set34 shows the absence of the second near neighbour peak for palladium, further reducing the values in the vicinity of 4 \AA, at the final time.
With the exception of the latter case, in which nanoparticle disorder at 10ns increases, we expect a rearrangement of palladium atoms in the previous cases, while for gold the rearrangement will be minimal.

To support the analysis on chemical ordering, we further analyzed the surface composition and chemical ordering by means of the mixing parameter.

\begin{figure}[ht!]
\centering
\includegraphics[width=4.7cm]{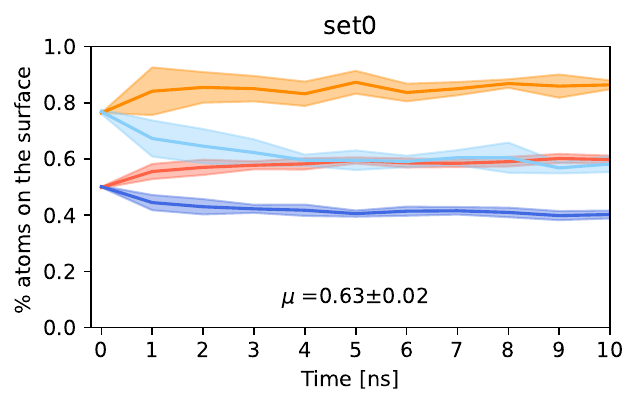}
\includegraphics[width=9.cm]{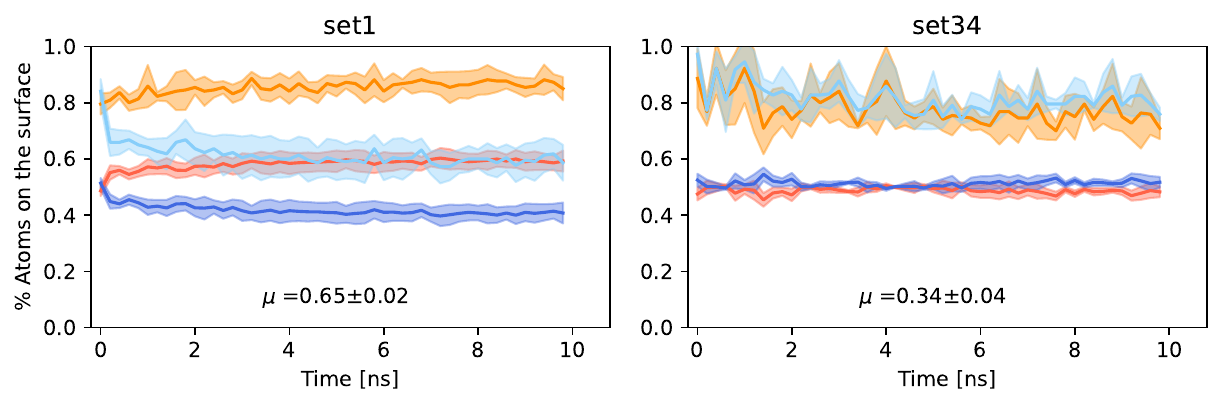}
\caption{Time-varying percentage of Au atoms (in red) and Pd atoms (in dark blue),
compared to the total number of surface atoms. Surface atoms were identified by their atop generalized coordination number. The shaded areas correspond to the variations on the simulations, while the solid line represents the average of the percentages. Also shown in light blue is the percentage of surface palladium atoms to the total number of palladium atoms and in orange the percentage of gold on the surface compared to the total number of atoms of gold. Each panel represents the sets1 and set34 as defined in Table \ref{fig:set}, and for each panel we also report the average mixing parameter at the final timestep (equal to 10 ns), with its standard deviation on the average of the independent simulations.}
\label{fig:PercMix400}
\end{figure}

We analyzed (i) the percentages of type A atoms on the surface compared to the total number of atoms on the surface (both type A and type B). Represented in red for gold and in dark blue for palladium in Figure \ref{fig:PercMix400};
(ii) the percentages of type A atoms on the surface compared to the total number of type A atoms. Represented in light blue for Pd, and in orange for gold in Fig. \ref{fig:PercMix400}.

For both, surface atoms were identified by their atop generalized coordination number. The panels show the average over the independent simulations of the percentages on the solid line and the respective standard deviation across the shaded areas.
We can see that for set34 both quantities are very similar for gold and palladium, indicating a comparable amount of palladium and gold atoms both on the surface and in the bulk, respectively.
In all other sets of implicit environment and of vacuum we observe that the amount of gold atoms on the surface is higher than both the amount of palladium atoms and the amount of gold atoms in the bulk. 
In fact, while surface gold atoms tend to be in most sets about 60 percent, palladium atoms are about 40 percent. Furthermore, generally 80 percent of the available gold atoms are arranged on the surface, while as for palladium it has about 60 percent atoms on the surface and the remaining 40 percent in the bulk.

We can then observe how the parameters $\rho$ and $\eta$ affect it. Varying $\rho$ while keeping the same $\eta$ we notice that in the most vacuum-like cases it exclusively changes the shape of the percentage curve a little bit: the amount of palladium on the surface decreases, as expected, since the surface energy of gold turns out to be less than that of palladium. 
So with the exception of set 34 we are not yet in a regime where the effect of the environment is such that it makes up the difference.
Looking then at the percentage of Pd atoms to the total number of Pd atoms (relative percentage) it is possible to appreciate slightly different trends. In fact, all the relative percentages of palladium possess drops, however, in some cases these stabilize quickly, while in others the process is slower. 

In addition, we can observe that the mixing parameter (whose mean and error are reported in of figure \ref{fig:PercMix400}) is lower for set34 than for the other sets, indicating a high segregation of the former. While for the other sets this shows us that the segregation is higher between gold and palladium.
By combining this information with that provided by trajectory analysis, we recognise a ball-cup chemical ordering.

\subsubsection{Coalescence of Au$_{55}$Pd$_{55}$ at 600 K}

Let us now comment the coalescence of Au$_{55}$ against a Pd$_{55}$ at 600K. Again we focus only on the coalescence of two icosahedral seeds.

\begin{figure}[ht!]
    \centering
    \includegraphics[width=2.55cm]{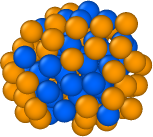}
    \includegraphics[width=2.55cm]{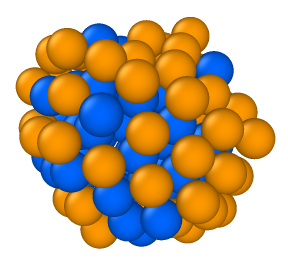}
    \includegraphics[width=2.55cm]{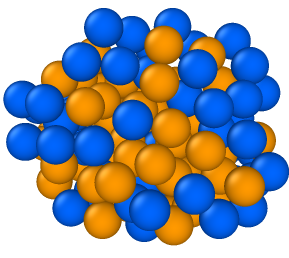}
    \caption{One snapshot for ${Au_{55}^{Ih}}$-${Pd_{55}^{Ih}}$ at 600K in vacuum (on the left) and in two sets of the environment: set1 and set34 (in center and right) at final time. Final time is equal to 10 ns.}
    \label{fig:snap_55_600K}
\end{figure}

From the qualitative analysis, it could be inferred that from a few nanoseconds to the final time, the nanoaggregate appears to be compact and well mixed on the surface. Unlike the case at 400K temperature, a Ball-Cup type pattern is not present here.
Furthermore, it is not possible to indicate a prevalence of five-fold symmetries, although these are sometimes present. Within the same set, as the independent simulation examined varied, they were in fact sometimes present and sometimes absent. Finally, we note that in the case where five-fold symmetry is present, it is not necessarily palladium atoms, or only palladium atoms, but gold also forms five-fold structures.
Figure \ref{fig:snap_55_600K} shows a snapshot for the vacuum (at left), in the middle the set1 (in center), and one for the set34 (right), which possess a particular behavior and serves as a reference also for the set33. 

The comparison between Figures  \ref{fig:snap_55_400K} and \ref{fig:snap_55_600K} reveals the effect of the external temperature. From the quantitative analysis we can observe that for all observed quantities the fluctuations of the averages, as well as the standard deviation, are higher than those of the simulations at 400K temperature.

\begin{figure}[ht!]
    \centering
    \includegraphics[width=9.cm]{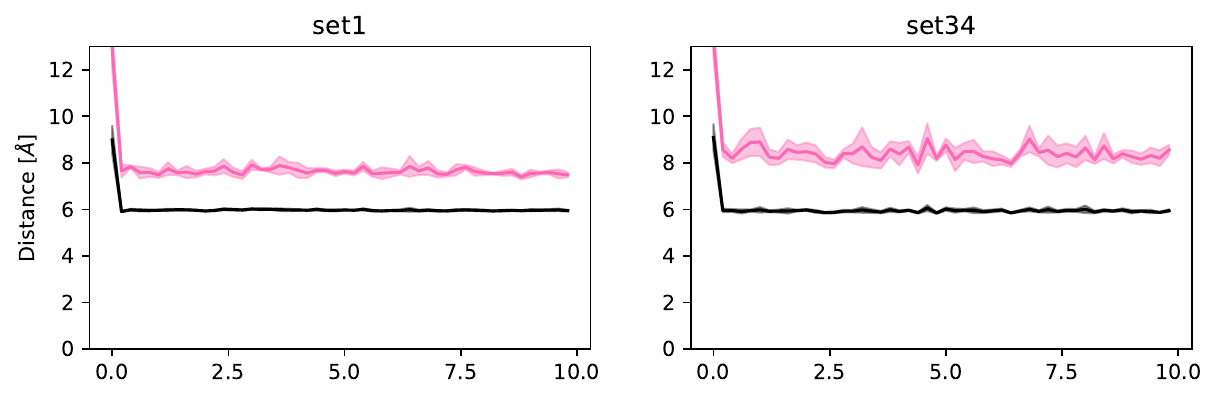}
    \caption{Time-varying radius of gyration of the two chemical species: in orange is represented the mean radius of gyration of Au and in blue that of palladium.}
    \label{fig:Unionenv600}
\end{figure}

\paragraph{Kinetic of the sintering process}
As previously from both radius of gyration and maximum pair distance (respectively black and pink lines in Fig. \ref{fig:Unionenv600}, reported here for the set1 and set34, we can observe that the structure compacts after a few nanoseconds. However, the greater oscillation of the maximum pair distance would lead us to assume rearrangements that go to slightly greater expansion and compaction of the nanoparticle. We emphasize that we are talking about oscillations anyway below 2 \AA~ in amplitude.

\begin{figure}[ht!]
    \centering
    \includegraphics[width=6cm]{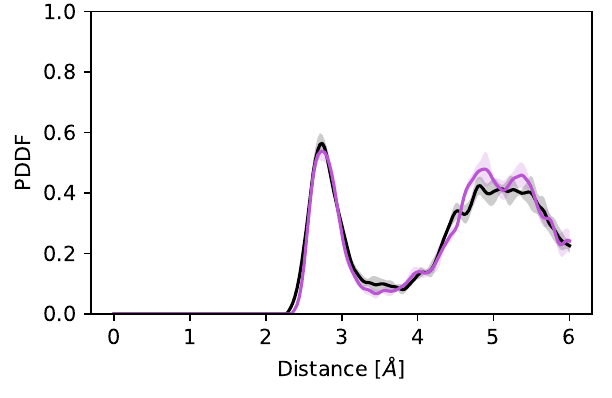}
    \caption{Pair distance distribution function mean of the whole cluster. In black is represented the PDDF at initial time (equal to the contact time) while in purple is represented the PDDF at final time. Final time is equal to 10 ns for all simulations.}
    \label{fig:PDFmean600}
\end{figure}

\paragraph{Structural characterization}
Trying to characterise the geometry of the nanoparticle, we can observe from the pair distance distribution function at the contact time (black line in the Figure \ref{fig:PDFmean600} ) and at the final time (purple line in the figure \ref{fig:PDFmean600} ), that the nanoparticle at the final time do not presents a more ordered structure. In fact, the peak of the neighbouring seconds, which is completely absent at the initial time, do not forms neither at the final time.
So we do not expected a final structure more ordered than the initial one.

The signature and pattern analysis of the common neighbour analysis was carried out for the sake of completeness. However, it is of little significance: due to the high disorder of the nanoparticle at 10ns at 600K, it was not possible to distinguish well-defined structures. As already anticipated from the trajectory analysis.

\paragraph{Chemical ordering and surface composition}

\begin{figure}[ht!]
    \centering
    \includegraphics[width=9.cm]{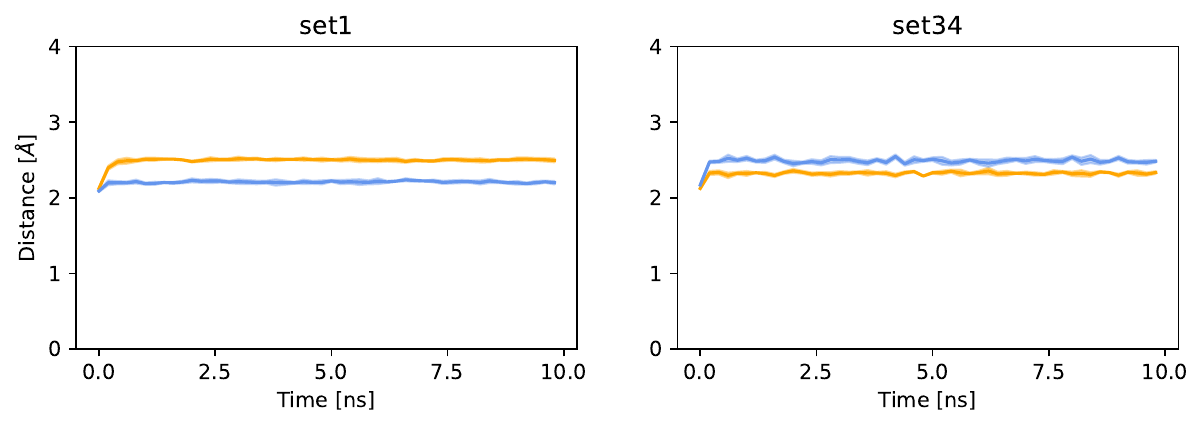}
    \caption{Time-varying radius of gyration of the two chemical species: in orange is represented the mean radius of gyration of Au and in blue that of palladium.}
    \label{fig:GYRAuPd600}
\end{figure}

Looking separately at the radius of gyration of gold and palladium (orange and blue lines in Fig. \ref{fig:GYRAuPd600} ), we can see that that of the latter is smaller, except for set34 where the trend is reversed and set33 where the sizes of the two are comparable, while in all other sets they behave similarly to set1.

Regarding PDDFs of the two chemical species, we can see that except for palladium of sets: set2, set3, set4, set15, even at the final time there is no peak of neighboring seconds neither for the overall PDDF of the nanoparticle, nor for the gold part alone, nor for the palladium part alone. The absence of a rearrangement even in the last chemical species mentioned constitutes a difference from all the cases simulated previously. This suggests the absence of a geometrically well-defined structure also for palladium.

\begin{figure}[ht!]
    \centering
    \includegraphics[width=9.cm]{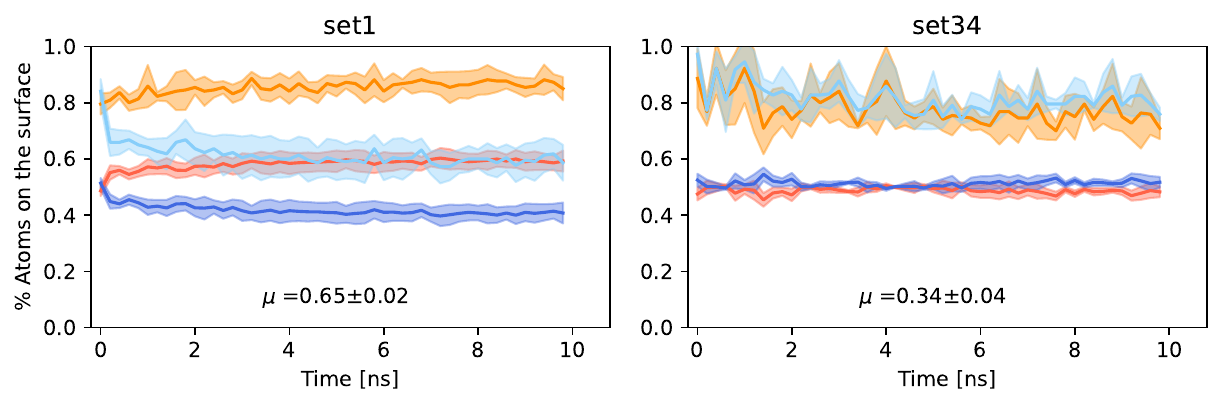}
    \caption{Time-varying percentage of Au atoms (in red) and Pd atoms (in dark blue), compared to the total number of surface atoms. Surface atoms were identified by their atop generalized coordination number. The shaded areas correspond to the variations on the simulations, while the solid line represents the average of the percentages. Also shown in light blue is the percentage of surface palladium atoms to the total number of palladium atoms and in orange the percentage of gold on the surface compared to the total number of atoms of gold. Each panel represents the sets defined in Table \ref{fig:set}, and for each panel we also report the average mixing parameter at the final timestep, with its standard deviation on the average of the independent simulations.}
    \label{fig:MeanPerc600}
\end{figure}

Figure \ref{fig:MeanPerc600} reports the total and relative percentages, respectively. 
We observe the total percentage (represented with darker lines in figure) behaves very similarly to the 400K temperature case, with about 60 percent gold atoms on the surface and 40 percent palladium.
We again consider as exceptions the set33 in which the two are nearly equivalent and the set34 in which the percentage of surface palladium is slightly greater than that of gold.
As for the relative percentage, however, it is very high (about 90 percent) for gold: indicating that almost all of the gold is on the surface. That for palladium (light blue lines in figure \ref{fig:MeanPerc600}), on the other hand, turns out to be very oscillating within the same set: undergoing variances of as much as 20 percent.
Similar to what has already been observed, for set33 and 34 there are special cases where the percentages of palladium on the surface are very high and comparable to those of gold.

As expected by the qualitative analysis, the mixing parameter reported in figure \ref{fig:MeanPerc600}, with its standard deviation, is lower than that at 400K temperature, with values between 0.35 and 0.5, confirming greater mixing of gold and palladium than in the previous case.

Let us therefore recall that at a temperature of 600K, gold could melt to such small nanoparticles. Therefore, we can deduce from the comparison between these simulations and those of the case at 400K that the melting process most probably interfered with the coalescence process, leading to more mixed and structurally disordered nanoalloys.


\subsection{The case of $Au_{561}^{Ih}$-$Pd_{561}^{Ih}$ }

After the study of the effect of a realistic implicit environment, we analysed the situation when two much larger seeds collides. In particular, we investigated the coalescence of one gold and one palladium icosahedra of 561 gold atoms. We limit our analysis to only one case simply because our focus is to investigate the effect of the implicit environment and under which conditions the compotion at the surface is latered. They were immersed in five different types of implicit environment, shown in light purple in Table \ref{fig:set}. 
In this case, the simulations were carried out at a temperature of 600K, however due to the larger size, the gold melting effect is not incisive.
All the magnitudes obtained are the averages over 4 independent simulations of the magnitudes of interest, each reported with its own standard deviation.
We emphasise that they constitute the large-scale analogue of some of the sets composed of 55 gold and palladium atoms; for ease of understanding, the name of the set coincides with that of the previous sets for small clusters, with the same values of the parameters $\rho$ and $\eta$.
In set1 the bond is covalent and gold and palladium interact in the same way with the environment, in set2 and set3 we varied the $\eta$ for palladium. While in set21 and set31 we varied the type of interaction of palladium with the environment by varying its $\rho$.

The simulations were performed at a temperature of 600K and have a duration of 100 ns for set1 and 20ns for the other sets.
For each set, four independent simulations were performed and then the results were averaged.

\begin{figure}[ht!]
    \centering
    \includegraphics[width=9.cm]{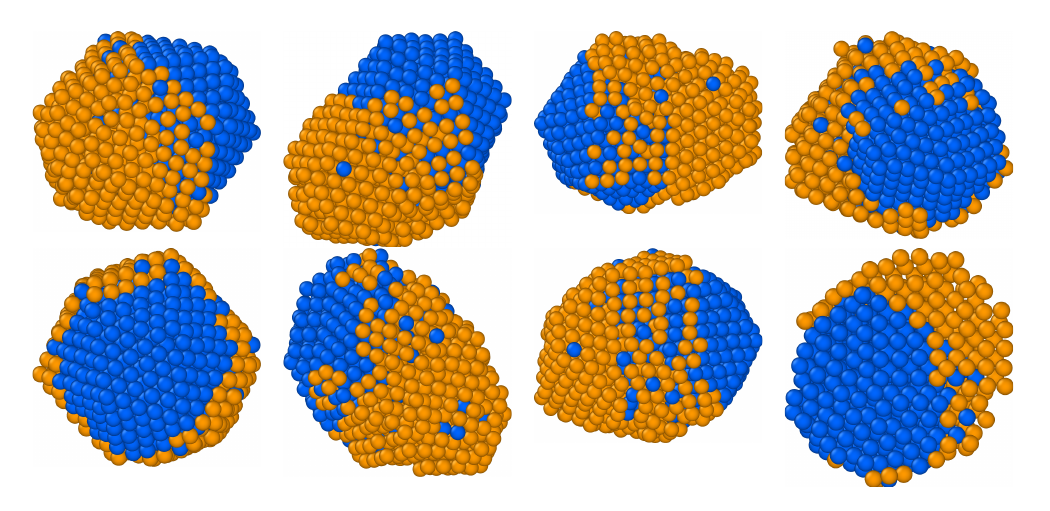}
    \caption{Snapshots of trajectories at final time}
    \label{fig:snapshot561}
\end{figure}

In Figure \ref{fig:snapshot561} we reported some snapshots at the final time.
Analysis of the trajectories shows that the nanoparticles are generally elongated. In all the simulations, we observed that the palladium tends to maintain five-fold symmetries and to be covered by the gold, which diffuses haphazardly above it, forming ball-cups and not shells shape.

\begin{figure}[ht!]
    \centering
    \includegraphics[width=15.cm]{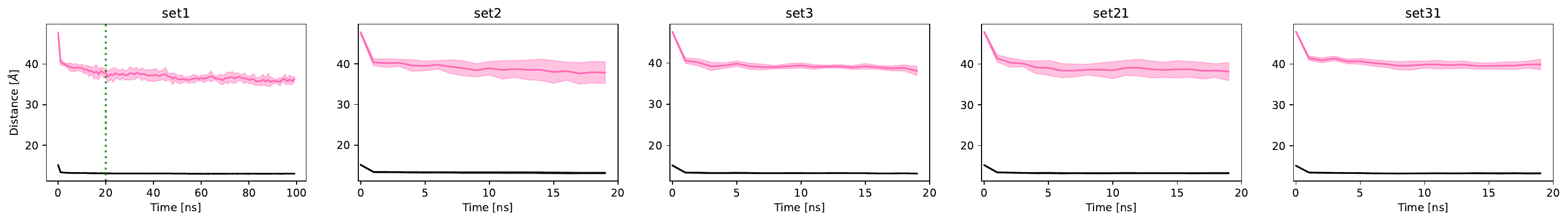}
    \caption{In black is represented the mean radius of gyration of whole cluster, in pink is the half maximum pair distance. All quantities reported are the average on the independent simulations: the shade represents their standard deviation. The vertical line represents the 20 ns.}
    \label{fig:Unionenv561}
\end{figure}

\paragraph{Kinetic of the sphericalisation process}
Although it decreases in the first nanoseconds, the maximum pair distance shown in pink in figure \ref{fig:Unionenv561} is clearly greater than the radius of the entire nanoparticle (shown in black), supporting the observation of nanoparticles that are not very compact and more elongated than in the case of small-sized nanoaggregates. 
We can also see that this is about 16  \AA higher than the analogous case in vacuum.

\begin{figure}
    \centering
    \includegraphics[width=15.cm]{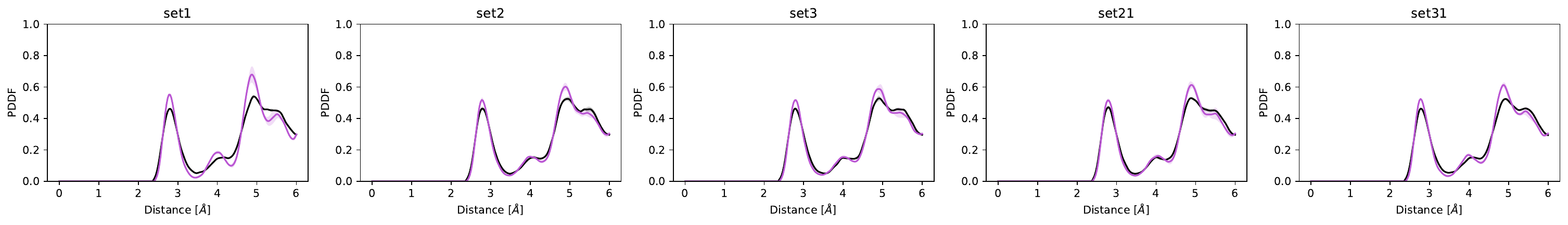}
    \caption{Pair distance distribution function mean of the whole cluster. In black is represented the PDDF at initial time (equal to the contact time) while in purple is represented the PDDF at final time. Final time is equal to 100 ns for set1 and equal to 20 ns for all the other simulations.}
    \label{fig:PDFmeanenv561}
\end{figure}

\paragraph{Structural characterization}
As clear in Fig.\ref{fig:PDFmeanenv561}, the pair distance distribution function at the final time shows a more pronounced peak for set 1, while it is better but with a minimal difference in the other sets. This difference between the sets can be attributed more to the different duration of the simulation than to the different parameters of the potential. In fact, for set1 the final PDDF depicted refers to 100 ns, while for the other sets to 20 ns.

We have also observe the presence for all the sets of the signatures (5,5,5), however with a tendency to decrease, although without cancelling completely for set1. We also observe a tendency for (4,2,1) to increase, up to around 40-60 per cent. The (4,2,2), on the other hand, tends to decrease by a few percentage points, stabilising at around 20 per cent. From the CNA patterns we can observe that only two independent simulations presents a (12(5,5,5)), which confirms the presence of icosahedral arrangements.

\begin{figure}[hb!]
    \centering
    \includegraphics[width=15.cm]{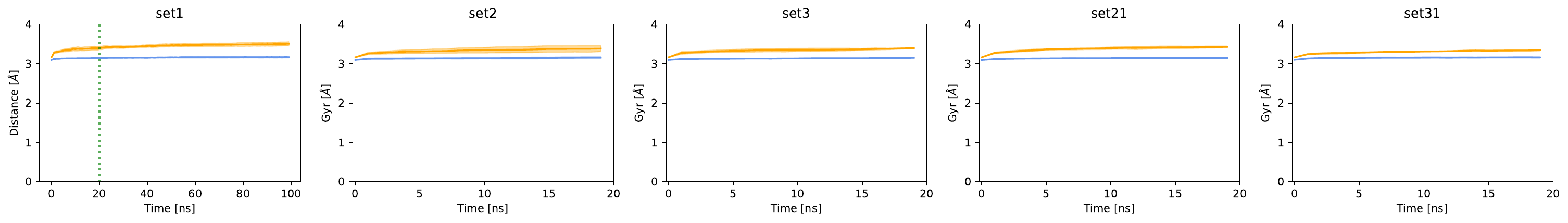}
    \caption{Time-varying radius of gyration of the two chemical species: in orange is represented the mean radius of gyration of Au and in blue that of palladium.}
    \label{fig:GYRAuPdenv561}
\end{figure}
\begin{figure}[hb!]
    \centering
    \includegraphics[width=15.cm]{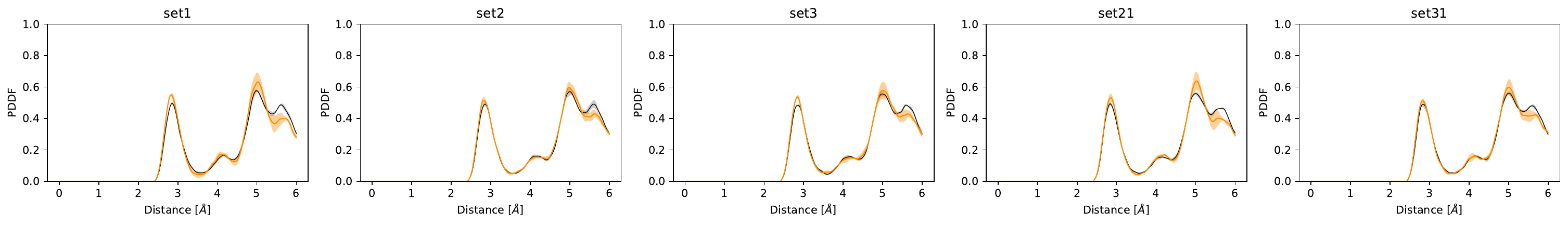}
    \caption{Pair distance distribution function mean of gold. In black is represented the PDDF at initial time (equal to the contact time) while in yellow is represented the PDDF at final time. Final time is equal to 100 ns for set1 and 20 ns for the others.
    }
    \label{fig:PDFmeanAuenv561}
\end{figure}

\paragraph{Chemical ordering and surface composition}

We notice that the radii of gyration for palladium (shown in light blue in figure \ref{fig:GYRAuPdenv561}) are smaller than those for gold (shown in orange in the same figure), thus suggesting that the former remains more compact around its centre of mass. 

\begin{figure}[ht!]
    \centering
    \includegraphics[width=15.cm]{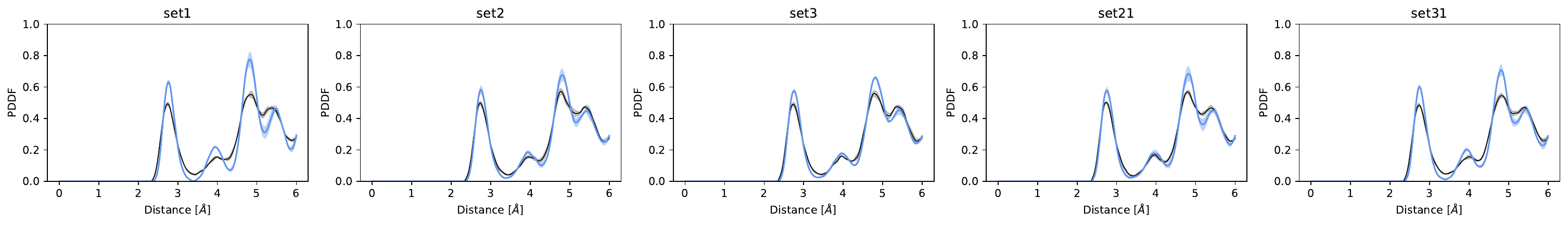}
    \caption{Pair distance distribution function mean of palladium. In black is represented the PDDF at initial time (equal to the contact time) while in blue is represented the PDDF at final time. Final time is equal to 100 ns for set1 and 20 ns for the others.}
    \label{fig:PDFmeanPdenv561}
\end{figure}

From the analysis of the PDDF (Fig. \ref{fig:PDFmeanAuenv561} and Fig. \ref{fig:PDFmeanPdenv561}) at the final time, we observe that the peak of the near seconds is more pronounced compared to the initial time for palladium, while it remains almost similar for gold. This suggests an improvement in the geometry of the system as far as palladium is concerned, in agreement with the findings of the trajectory analysis.
We also observe for set1 (represented on the left panel), whose final PDDF is reported at a time of 100ns, the peak of the second neighbours is clearly better defined, suggesting that a greater rearrangement in the nanoparticle can be achieved by increasing the simulation time.

\begin{figure}[ht!]
    \centering
    \includegraphics[width=15.cm]{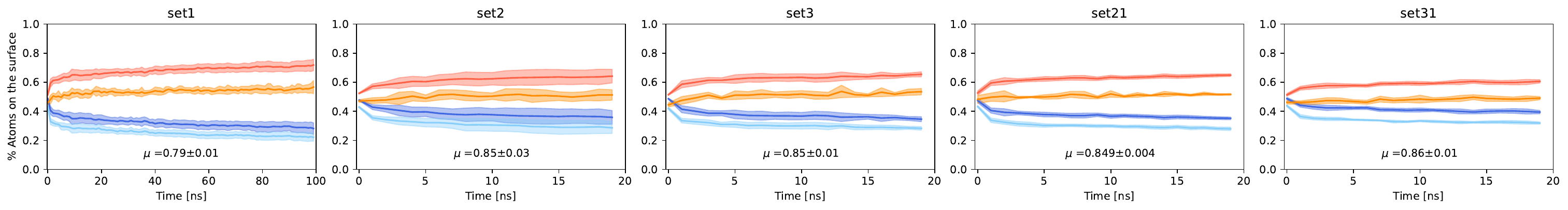}
    \caption{Time-varying percentage of Au atoms (in red) and Pd atoms (in dark blue), compared to the total number of surface atoms. Surface atoms were identified by their atop generalized coordination number. The shaded areas correspond to the variations on the simulations, while the solid line represents the average of the percentages. Also shown in light blue is the percentage of surface palladium atoms to the total number of palladium atoms and in orange the percentage of gold on the surface compared to the total number of atoms of gold. Each panel represents the chosen sets. We report the average mixing parameter at the final timestep, with its standard deviation on the average of the independent simulations.}
    \label{fig:MeanPercenv561}
\end{figure}

From the analysis of the surface compositions as depicted in Figure \ref{fig:MeanPercenv561}, we note that the number surface gold and palladium atoms tends to be constant as in the case of small clusters, with about 60 per cent surface gold and about 40 per cent palladium, which tends to increase slightly in the case of the final time at 100ns. On the other hand, the relative percentage decreases significantly for both chemical species: gold has about 45 per cent surface atoms, while palladium has 30 per cent.
The mixing parameter mirrors what is observed qualitatively, settling for all sets around 0.85: indicating a trend towards the presence of a fully segregated nanoalloy in its components: gold and palladium. Thus suggesting the presence of a ball-cup with the two chemical species highly segregated.

\section*{Conclusions}
Among the variety of nanoalloys, gold and palladium exhibit superior catalytic activity and enhanced selectivity as compared to their mono-metallic counterparts due to synergistic effects that have been widely investigated in both heterogeneous thermal catalysis and electrocatalysis.
Through classical Molecular Dynamics simulations we modelled the coalescence of gold and palladium nanoaprticles both in vacuum, and adding a metal-environment interaction potential. The latter reproduced via the Cortes-Huerto's formulation that enables to select the type of interaction (covalent, pairwise or strongly interacting) and its strength.
We considered the sintering of small sized seeds leading to a nanoparticle of 110 atoms, simulated at of 400K and 600K;  and the sintering of two icosahedra of 561 atoms simulated only at 600K. In both scenarios, we selected icosahedral seeds of same size, hence considering only a 50\% chemical composition.
We selected the parameters of the implicit environment in such a way that gold-environment is always a covalent like interactions. On the other hand, we varied the interaction of palladium atoms with the surroundings from (i) covalent (soft and strong); (ii) pairwise and (iii) strongly interacting.
Furthermore, we varied the relative weight of the interaction with the environment. We first selected the case where gold and palladium have the same binding, therefore, we kept the binding to gold fixed and varied the binding to palladium, increasing the interaction of palladium with the environment. 
For completeness, we also investigated a case in which the interaction of palladium with the environment was repulsive.

For all configurations, the disappearance of the neck leading to an almost spherical nanoparticle occurs within the first nanosecond. At 600~K, we should note that the 55-atoms Au-seed is almost melted affecting the overall coalescence process, leading to nanoparticles with a kind of mixed behaviour.
The simulations at 400~K for the coalescence of 55-atoms seeds and those arising from the sintering of two 561-atoms at 600~K show a similar behavior with the same interaction with the selected environment. We observe a strong  tendency to form ball-cup structures, in agreement with global optimisation results, with Pd atoms rearranging in the innermost layers, and Au atoms to diffuse above Pd.
If the environment interacts strongly with Pd and less with Au interacting we observe a higher amount of palladium on the surface and eventually the formation of a Janus-like pattern.  
We do believe that the effect of an interacting environment is therefore important to reveal the physico-chemical properties of Au-based nanoalloys, mainly used for nanoanteannas and plasmonic devices.

\section*{Author Contributions}
SZ performed all the simulations and FB had the initial idea. Both wrote the paper.

\section*{Conflicts of interest}
There are no conflicts to declare.

\printbibliography


\end{document}